\documentclass{appolb}
\usepackage{graphicx}
\usepackage{subfigure}

\begin{document}
\title{The lattice Landau gauge gluon propagator at zero and finite temperature \thanks{Presented by Paulo J. Silva at the International Meeting ``Excited QCD'', Peniche, Portugal, 6 - 12 May, 2012.}%
}
\author{Orlando Oliveira, Paulo J. Silva
\address{Centro de F\'{i}sica Computacional, Universidade de Coimbra, 3004-516 Coimbra, Portugal}
}
\maketitle
\begin{abstract}
We study the Landau gauge gluon propagator at zero and finite temperature using lattice simulations. Particular attention is given to the finite size effects and to the infrared behaviour. 
\end{abstract}
\PACS{11.15.Ha; 12.38.Gc}
  
\section{The gluon propagator at zero temperature}

In lattice QCD, the finite lattice spacing and finite lattice volume 
effects on the gluon propagator can be investigated with the help of 
lattice simulations at several lattice spacings and physical volumes. 
Here we report on such a calculation.
For details on the lattice setup see \cite{OliSi12}. 

\begin{figure}[t]
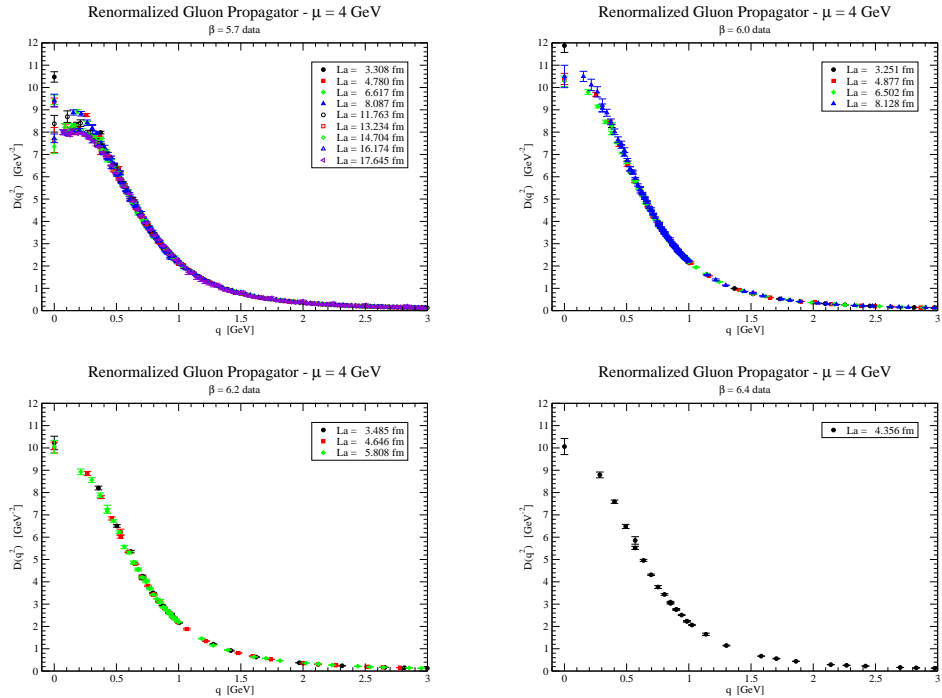
 
   \centering
   \subfigure{ \includegraphics[scale=0.235]{figures/glue_R4GeV_B5.7.eps} } \qquad
   \subfigure{ \includegraphics[scale=0.235]{figures/glue_R4GeV_B6.0.eps} }

   \subfigure{ \includegraphics[scale=0.235]{figures/glue_R4GeV_B6.2.eps} } \qquad
   \subfigure{ \includegraphics[scale=0.235]{figures/glue_R4GeV_B6.4.eps} }
  \caption{Renormalized gluon propagator for $\mu = 4$ GeV for all lattice simulations.}
   \label{fig:gluevol}
\end{figure}

In figure \ref{fig:gluevol}, we show the renormalized gluon propagator at $\mu = 4$ GeV for all lattice simulations. Note that we compare our data with the large volume simulations performed by the Berlin-Moscow-Adelaide collaboration \cite{BMA09} -- see \cite{OliSi12} for details. In each plot we show data for a given value of $\beta$, i.e. data in the same plot has the same lattice spacing. The plots show that, for a given lattice spacing, the infrared gluon propagator decreases as the lattice volume increases. For larger momenta, the lattice data is less dependent on the lattice volume; indeed, for momenta above $\sim$900 MeV the lattice data define a unique curve. 

\begin{figure}[t]
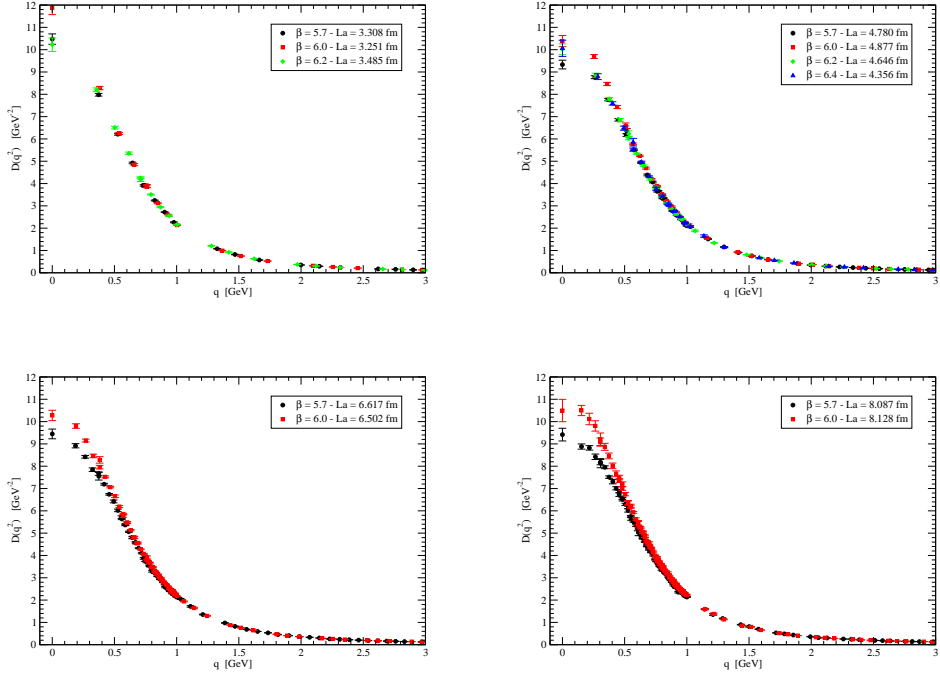
 
   \centering
   \subfigure{ \includegraphics[scale=0.235]{figures/glue_R4GeV_V3.3fm.eps} } \qquad
   \subfigure{ \includegraphics[scale=0.235]{figures/glue_R4GeV_V4.6fm.eps} }

\vspace{0.6cm}
   \subfigure{ \includegraphics[scale=0.235]{figures/glue_R4GeV_V6.6fm.eps} } \qquad
   \subfigure{ \includegraphics[scale=0.235]{figures/glue_R4GeV_V8.1fm.eps} }
  \caption{Comparing the renormalized gluon propagator at $\mu = 4$ GeV for various lattice spacings and similar physical volumes.}
   \label{fig:gluespac}
\end{figure}

We can also investigate finite volume effects by comparing the renormalized gluon propagator computed using the same physical volume but different $\beta$ values. We are able to consider 4 different sets with similar physical volumes --- see figure \ref{fig:gluespac}. Although the physical volumes considered do not match perfectly, one can see in figure \ref{fig:gluespac} that for momenta above $\sim$ 900 MeV the lattice data define a unique curve. This means that the renormalization procedure has been able to remove all dependence on the ultraviolet cut-off $a$ for the mid and high momentum regions. However, a comparison between figures \ref{fig:gluevol} and \ref{fig:gluespac} shows that, in the infrared region, the corrections due to the finite lattice spacing seem to be larger than the corrections associated with the finite lattice volume. In particular, figure  \ref{fig:gluespac} shows that the simulations performed with $\beta=5.7$, i.e., with a coarse lattice spacing, underestimate the gluon propagator in the infrared region. In this sense, the large volume simulations performed by the Berlin-Moscow-Adelaide collaboration provide a lower bound for the continuum infrared propagator.

\section{The gluon propagator at finite temperature}

We also aim to study how temperature changes the gluon propagator. At finite temperature, the gluon propagator is described by two tensor structures, 

\begin{equation}
D^{ab}_{\mu\nu}(q)=\delta^{ab}\left(P^{T}_{\mu\nu} D_{T}(q_4,\vec{q})+P^{L}_{\mu\nu} D_{L}(q_4,\vec{q}) \right) \nonumber
\label{tens-struct}
\end{equation}
where the transverse and longitudinal projectors are defined by
\begin{equation}
P^{T}_{\mu\nu} = (1-\delta_{\mu 4})(1-\delta_{\nu 4})\left(\delta_{\mu \nu}-\frac{q_\mu q_\nu}{\vec{q}^2}\right) \quad , \quad
P^{L}_{\mu\nu} = \left(\delta_{\mu \nu}-\frac{q_\mu q_\nu}{{q}^2}\right) - P^{T}_{\mu\nu} \, ;
\label{long-proj}
\end{equation}
the transverse $D_T$ and longitudinal  $D_L$ propagators are given by
\begin{equation}
D_T(q)=\frac{1}{2V(N_c^2-1)}\left(\langle A_i^a(q) A_i^a(-q)\rangle-\frac{q_4^2}{\vec{q}^2} \langle A_4^a(q) A_4^a(-q)\rangle \right) \nonumber
\end{equation}

\begin{equation}
D_L(q)=\frac{1}{V(N_c^2-1)}\left(1+\frac{q_4^2}{\vec{q}^2} \langle A_4^a(q) A_4^a(-q)\rangle\right) \nonumber
\end{equation}

On the lattice, finite temperature is introduced by reducing the temporal extent of the lattice, i.e. we work with lattices $L_s^3 \times L_t$, with $L_t \ll L_s$. The temperature is defined by $T=1/a L_t$.

 In table \ref{tempsetup} we show the lattice setup of our simulation. Simulations in this section have been performed with the help of Chroma library \cite{chroma}. For the determination of the lattice spacing we fit the string tension data in \cite{bali92} in order to have a function $a(\beta)$. Note also that we have been careful in the choice of the parameters, in particular we have only two different spatial physical volumes: $\sim(3.3\mbox{fm})^3$ and $\sim(6.5\mbox{fm})^3$. This allows for a better control of finite size effects.

\begin{table}
\begin{center}
\begin{tabular}{cccccc}
\hline
Temp. (MeV) &	$\beta$ & $L_s$ &  $L_t$ & a [fm] & 1/a (GeV) \\
\hline
121 &	6.0000 & 32,64 & 	16 & 	0.1016 &  	1.9426 \\
162 &	6.0000 & 32,64 & 	12 & 	0.1016 & 	1.9426 \\
243 &	6.0000 & 32,64 & 	8 & 	0.1016 & 	1.9426 \\
260 &	6.0347 & 68    & 	8 & 	0.09502 & 	2.0767 \\
265 &	5.8876 & 52    & 	6 & 	0.1243 & 	1.5881 \\
275 &	6.0684 & 72    & 	8 & 	0.08974 & 	2.1989 \\
285 &	5.9266 & 56    & 	6 & 	0.1154 & 	1.7103 \\
290 &	6.1009 & 76    & 	8 & 	0.08502 & 	2.3211 \\
305 &	5.9640 & 60    & 	6 & 	0.1077	 &      1.8324 \\
305 &	6.1326 & 80    & 	8 & 	0.08077 & 	2.4432 \\
324 &	6.0000 & 32,64 & 	6 & 	0.1016	 &      1.9426 \\
486 &	6.0000 & 32,64 & 	4 & 	0.1016	 &      1.9426 \\
\hline
\end{tabular}
\end{center}
\label{tempsetup}
\caption{Lattice setup used for the computation of the gluon propagator at finite temperature.}
\end{table}

\begin{figure}[t]
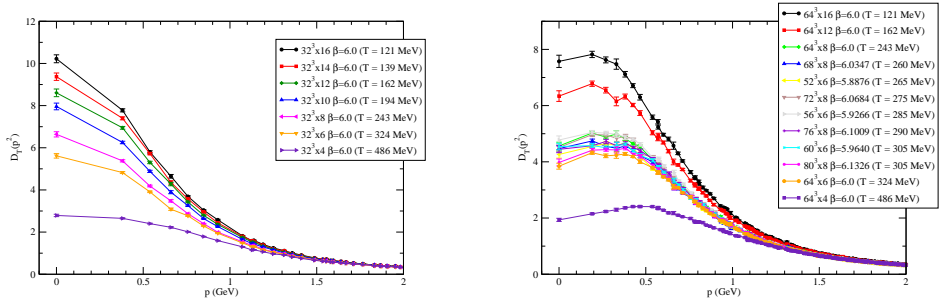
 
   \centering
   \subfigure{ \includegraphics[scale=0.222]{figtemp/trans32.eps} } \qquad
   \subfigure{ \includegraphics[scale=0.222]{figtemp/trans64.eps} }
  \caption{Transverse gluon propagator for  $\sim(3.3\mbox{fm})^3$ (left) and $\sim(6.5\mbox{fm})^3$ (right) spatial lattice volumes.}
   \label{fig:transtemp}
\end{figure}

\begin{figure}[t]
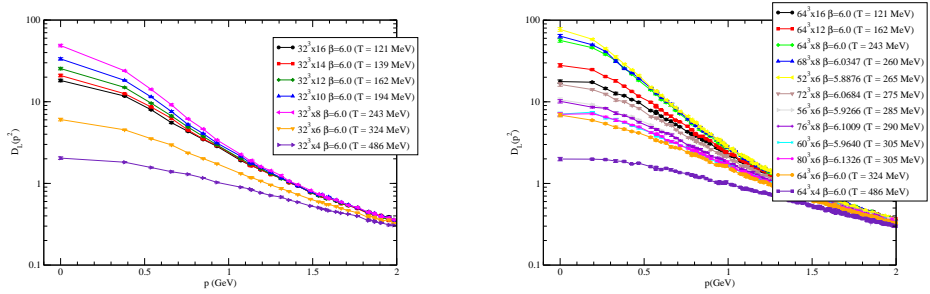
 
   \centering
   \subfigure{ \includegraphics[scale=0.215]{figtemp/long32.eps} } \qquad
   \subfigure{ \includegraphics[scale=0.215]{figtemp/long64.eps} }
  \caption{Longitudinal gluon propagator for  $\sim(3.3\mbox{fm})^3$ (left) and $\sim(6.5\mbox{fm})^3$ (right) spatial lattice volumes.}
   \label{fig:longtemp}
\end{figure}

\begin{figure}[t] 
   \centering
   \subfigure{ \includegraphics[scale=0.25]{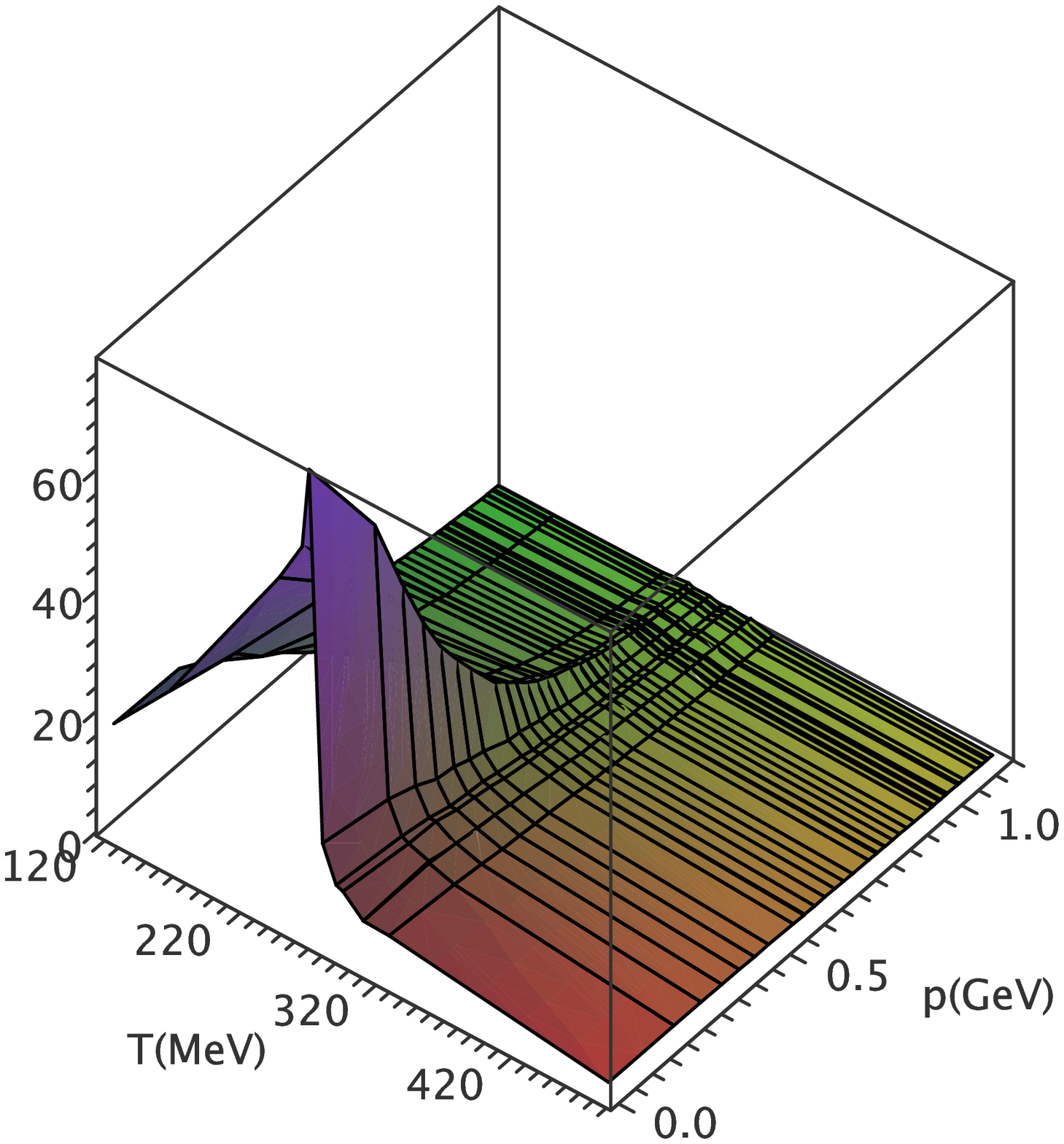} } \qquad
   \subfigure{ \includegraphics[scale=0.25]{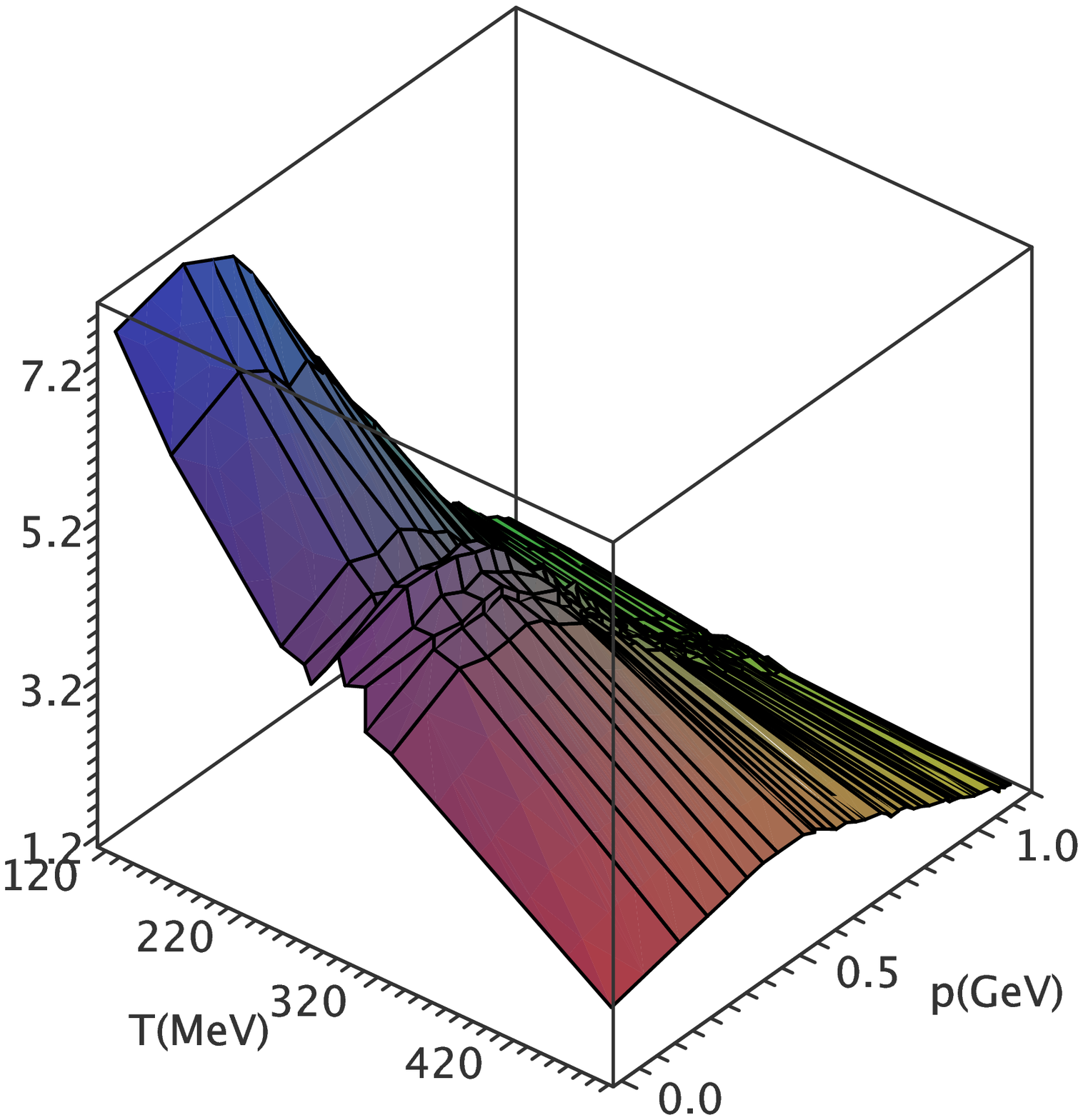} }
  \caption{Longitudinal (left) and transverse (right) gluon propagator as a function of momentum and temperature for a $\sim(6.5\mbox{fm})^3$ spatial lattice volume.}
   \label{fig:3dtemp}
\end{figure}

Figures \ref{fig:transtemp} and \ref{fig:longtemp} show the results obtained up to date. We see that the transverse propagator, in the infrared region, decreases with the temperature. Moreover, this component shows finite volume effects; in particular, the large volume data exhibits a turnover in the infrared, not seen at the small volume data. The longitudinal component increases for temperatures below $T_c\sim 270\, \mbox{MeV}$. Then the data exhibits a discontinuity around $T_c$, and the propagator decreases for $T > T_c$. The behaviour of the gluon propagator as a function of the temperature can also be seen in the 3d plots shown in figure \ref{fig:3dtemp}.

\begin{figure}[h]
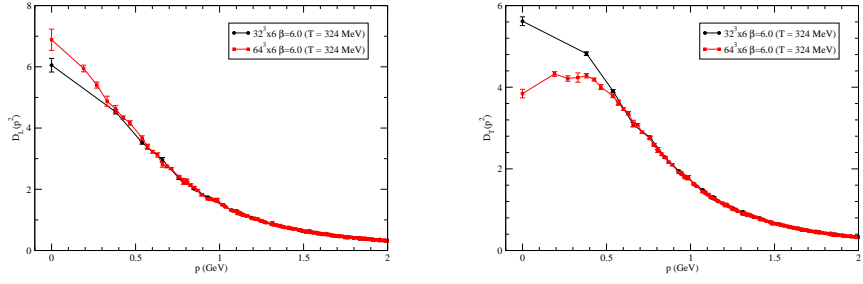
 
   \centering
   \subfigure{ \includegraphics[scale=0.215]{figtemp/long324MeV.eps} } \qquad
   \subfigure{ \includegraphics[scale=0.215]{figtemp/trans324MeV.eps} }
  \caption{Longitudinal (left) and transverse (right) gluon propagator for different spatial lattice volumes at T=324 MeV.}
   \label{fig:finvoltemp}
\end{figure}

\begin{figure}[t]
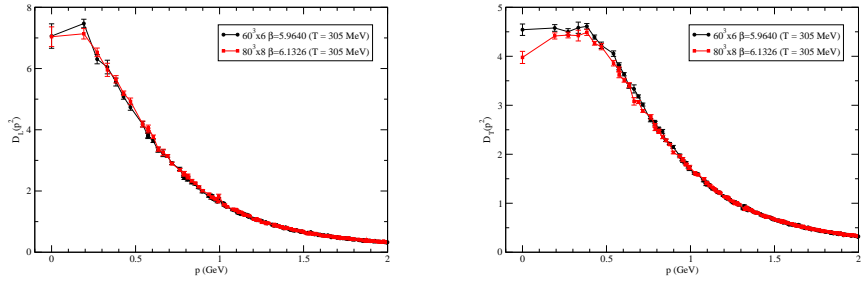
 
   \centering
   \subfigure{ \includegraphics[scale=0.215]{figtemp/long305MeV.eps} } \qquad
   \subfigure{ \includegraphics[scale=0.215]{figtemp/trans305MeV.eps} }
  \caption{Longitudinal (left) and transverse (right) gluon propagator for different lattice spacings (but similar physical volume) at T=305 MeV.}
   \label{fig:lattspactemp}
\end{figure}

As shown above, data for different physical (spatial) volumes exhibits finite volume effects. This can be seen in more detail in figure \ref{fig:finvoltemp}, where we show the propagators for two volumes at T=324 MeV. Moreover, we are also able to check for finite lattice spacing effects at T=305 MeV, where we worked out two different simulations with similar physical volumes and temperatures, but different lattice spacings. For this case, it seems that finite lattice spacing effects are under control, with the exception of the zero momentum for the transverse component -- see figure \ref{fig:lattspactemp}.

Our results show that a better understanding of lattice effects is needed before our ultimate goal, which is the modelling of the propagators as a function of momentum and temperature.

\section*{Acknowledgments}
Paulo Silva is supported by FCT under contract SFRH/BPD/40998/2007. Work supported by projects CERN/FP/123612/2011, CERN/FP/123620/2011 and PTDC/FIS/100968/2008, projects developed under initiative QREN financed by UE/FEDER through Programme COMPETE.

\end{document}